\documentclass[journal=jacsat,manuscript=article]{achemso}
\setkeys{acs}{articletitle = true}

\usepackage[version=3]{mhchem} 

\usepackage{amsmath}
\usepackage{amsfonts}
\usepackage{lscape}  
\usepackage{rotating} 
\usepackage{siunitx}
\usepackage{multirow}
\usepackage{multicol}
\usepackage[utf8]{inputenc}
\usepackage{tablefootnote}
\usepackage{color}
\usepackage[ruled]{algorithm2e}
\usepackage{booktabs}
\usepackage{xfrac}
\usepackage[flushleft]{threeparttable}
\usepackage{xcolor}
\newcommand{\rev}[1]{\textcolor{black}{#1}}
\newcommand{\revnew}[1]{\textcolor{black}{#1}}

\newcommand{\etal}{\textit{et al.}}

\newcommand{\bld}{\boldsymbol}
\newcommand{\mrm}{\mathrm}
\newcommand{\ket}[1]{\vert #1 \rangle}

\newcommand{\bra}[1]{\langle #1 \vert}

\newcommand{\Dbraket}[2]{\langle #1 \hspace{.10em} \vert \hspace{.10em}  #2 \rangle}

\newcommand{\Tbraket}[3]{\langle #1 \hspace{.10em} \vert \hspace{.10em} #2 \hspace{.10em} \vert \hspace{.10em} #3 \rangle}

\author{Sarai Dery Folkestad}
\affiliation{Department of Chemistry, Norwegian University of Science and Technology, N-7491 Trondheim, Norway}
\author{Henrik Koch}
\affiliation{Scuola Normale Superiore, Piazza dei Cavaleri 7, 56126 Pisa, Italy}
\alsoaffiliation{Department of Chemistry, Norwegian University of Science and Technology, N-7491 Trondheim, Norway}
\email{henrik.koch@sns.it}  

\title{Equation-of-motion MLCCSD and CCSD-in-HF oscillator strengths and their application to core excitations}

\abbreviations{IR,NMR,UV}

\begin{document}


\begin{abstract}
We present an implementation of equation-of-motion oscillator strengths for the multilevel CCSD (MLCCSD) model where CCS is used as the lower level method (CCS/CCSD).  
In this model, the double excitations of the cluster operator are restricted to an active orbital space, whereas the single excitations are unrestricted.
Calculated nitrogen K-edge spectra of adenosine, adenosine triphosphate (ATP), and an ATP-water system are used to demonstrate the performance of the model.
Projected atomic orbitals (PAOs) are used to partition the virtual space into active and inactive orbital sets. Cholesky decomposition of the Hartree-Fock density is used to partition the occupied orbitals. This Cholesky-PAO partitioning is cheap, scaling as $\mathcal{O}(N^3)$, and is suitable for the calculation of core excitations which are localized in character.
By restricting the single excitations of the cluster operator to the active space, as well as the double excitations, the CCSD-in-HF model is obtained. A comparison of the two models---MLCCSD and CCSD-in-HF---is presented for the core excitation spectra of the adenosine and ATP systems.
\end{abstract}

\section{Introduction}

The multilevel coupled cluster (MLCC) approach can be used to calculate excitation energies of molecular systems that are too large for the standard coupled cluster models.
In MLCC, the higher order excitations that are included in the cluster operator are restricted to an active orbital space. 
One can view the approach as applying a higher level of coupled cluster theory to the active orbital\rev{s}. 
The MLCC approach was introduced by Myhre \etal\citep{myhre2013extended,myhre2014mlcc,myhre2016multilevel} but is similar to the active space approach which has resulted from the multireference coupled cluster method of Oliphant and Adamowicz.\citep{oliphant1991multireference,piecuch1993state,kohn2006coupled,rolik2011cost}

In the multilevel coupled cluster singles and doubles (MLCCSD) model,\citep{myhre2013extended,myhre2014mlcc} CCSD\citep{Purvis1982} is applied to the active orbital space. Coupled cluster singles (CCS) and/or singles and perturbative doubles\citep{Christiansen1995} (CC2) is used for the inactive orbital space. 
With carefully selected active orbitals,
excitation energies of CCSD quality are obtained. When the active orbital space is enlarged, 
the MLCCSD excitation energies converge smoothly towards the CCSD excitation energies.
The MLCCSD oscillator strengths, within the coupled cluster response formalism,\citep{koch1990coupled,pedersen1997coupled} were reported in Ref.\citenum{myhre2016near}. However, these proof-of-concept calculations did not exploit the computational reductions offered by the multilevel framework and were performed using a standard CCSD code. 

Recently, we have reformulated and implemented the MLCCSD ground and excited state equations.\citep{folkestad2019multilevel}
We have found it sufficient to use CCS as the lower level model to obtain accurate valence excitation energies. This CCS/CCSD model is cheaper and simpler than the CC2/CCSD and CCS/CC2/CCSD models. Moreover, the CCS/CCSD model is compatible with properties derived within the equation-of-motion\citep{sekino1984linear,geertsen1989equation,bartlett2012coupled,krylov2008equation} (EOM) framework, as well as with coupled cluster response theory.
This is because the only modification with respect to CCSD is a restriction of the doubles part of the cluster operator to the active orbital space.
\rev{T}he CCS/CCSD model \rev{has been} used to calculate valence excitation energies for a system with \rev{more than fifty second row atoms}
and the computational scaling approaches that of CCS for sufficiently large inactive spaces.\citep{folkestadReduced}
In this paper, we present an implementation of EOM oscillator strengths\citep{stanton1993equation} for the CCS/CCSD model.

The success of an MLCC calculation relies heavily on the choice of the active orbital space. Two strategies are used to obtain the active orbitals: either information from a cheaper electronic structure model is used or localized (or semilocalized) orbitals in a subregion of the molecular system defines the active orbital space. The success of the first strategy relies on the accuracy of the cheaper electronic structure model. The use of correlated natural transition orbitals\citep{hoyvik2017,folkestad2019multilevel} (CNTOs) \rev{to determine the active space} is an example of such an approach. \rev{The CNTOs are similar to natural transition orbitals\citep{luzanov1976application,martin2003natural} (NTOs),  which are extensively used for analysis\citep{dreuw2005single,plasser2014new} and in reduced cost methods for excited states,\citep{mata2011incremental,helmich2011local,helmich2013pair,baudin2016lofex,baudin2017lofex} but are defined by using excitation vectors that are parameterized with both single and double substitutions with respect to the reference determinant.}

When an electronic excitation is localized in a region of the molecule, localized or semilocalized Hartree-Fock orbitals can be used to determine the active space.
Cholesky orbitals have been used in MLCC calculations for both core and valence excitation energies.\citep{myhre2014mlcc,myhre2016multilevel,myhre2016near}
Occupied Cholesky orbitals can be obtained through a partial, limited Cholesky decomposition of the idempotent Hartree-Fock density in the atomic orbital (AO) basis.\citep{aquilante2006fast,sanchez2010cholesky}
Virtual Cholesky orbitals can be obtained in the same way by considering the virtual Hartree-Fock density. 
This localization scheme is non-iterative and has cubic scaling with respect to system size.
Another option \rev{to determine the active virtual orbitals}, 
which can be used in conjunction with occupied Cholesky orbitals, 
are the projected atomic orbitals (PAOs).
PAOs have been used extensively in reduced cost electronic structure methods.\citep{pulay1983,saebo1993,hampel1996,schutz2001,korona2003local,kats2006local,kats2009multistate} The construction of PAOs is also a non-iterative procedure \rev{with} cubic scaling. 

In MLCCSD, the double excitations of the cluster operator are restricted to the active orbital space.
By also restricting the single excitations, 
we obtain a reduced space CCSD approach.
There are several reduced space coupled cluster approaches, such as the frozen core approximation, the frozen natural orbital approaches,\citep{taube2008frozen,landau2010frozen,deprince2013accurate,deprince2013accuracy,kumar2017frozen,mester2017reduced} and
the LoFEx\citep{baudin2016lofex,baudin2017lofex} and CorNFLEx\citep{baudin2017correlated} methods. 
The LoFEx and CorNFLEx methods are specialized for the calculation of accurate excitation energies. A truncated set of molecular orbitals (MOs) is determined by considering the dominant NTOs or CNTOs, obtained from a cheaper electronic structure method, and localized orbitals that overlap with these dominant NTOs/CNTOs. 
The reduced orbital space is increased until the excitation energy is converged to within a predefined threshold. As NTOs/CNTOs from a single excited state are used to determine the MOs that enter the coupled cluster calculation, LoFEx and CorNFLEx are state specific methods; the reduced space differs depending on the excited state.

In this work, we consider a reduced space CCSD approach (CCSD-in-HF) where Cholesky occupied orbitals and PAOs are used to obtain the active orbital space for a region of interest. 
Several excited states can be treated using the same truncated set of molecular orbitals, as long as the excitation processes are located in the region of interest. 
Preliminary studies using CC-in-HF to describe valence excitations have been reported.\citep{sanchez2010cholesky,eTpaper,folkestadReduced}
An iterative procedure, where the active space is increased and the excitation energies recomputed until convergence, as is done in LoFEx/CorNFLEx, is possible but has not yet been implemented.

In near edge X-ray absorption fine structure (NEXAFS) spectroscopy,\citep{stohr2013nexafs} a core electron is excited. Since the binding energy of a core electron is unique to a given atomic number, specific energy ranges correspond to the K-edge NEXAFS spectrum for the different atoms. 
The excitation energies are sensitive to the environment of the core excited atom and NEXAFS spectra can be used to probe the local environment. 
Because of the strong interaction between the core hole and the excited electron, core excitation processes are generally localized in character. 

With the development of the liquid microjet technique, studies of solutions and liquids with NEXAFS can  be performed routinely.
For reviews of the liquid microjet technique in soft X-ray spectroscopies, we refer the reader to Refs. \citenum{lange2013electronic} and \citenum{smith2017soft}.
Proper interpretation of NEXAFS spectra relies on accurate theoretical modeling. 
While it can be challenging to accurately model the NEXAFS spectra of small molecules \textit{in vacuo}, it is significantly more complicated for complex systems such as solutions and liquids.

The coupled cluster hierarchy of models can be used to accurately calculate core excitations, for instance by use of the core-valence separation\citep{cederbaum1980many} (CVS) approach of Coriani and Koch.\citep{coriani2015communication,coriani2016erratum}
Typical errors of CCSD core excitation energies, obtained within the CVS approximation, are on the order of $\SI{1}{\eV}$.
The errors can be significantly reduced by including triple excitations.\citep{myhre2018theoretical,myhre2019x,matthews2020eom}
Intensities can be obtained from coupled cluster linear response theory or from EOM coupled cluster theory. 
Myhre \etal\citep{myhre2016near} calculated the MLCCSD NEXAFS spectra at the carbon and oxygen edge for ethanal, propenal, and butanal, demonstrating excellent agreement with the CCSD spectra.
Their work showed that the multilevel coupled cluster models, using localized orbitals to determine the active space, is appropriate for the description of core excitation processes. While illustrating the usefulness of the MLCCSD model, this implementation was, as mentioned previously, not optimal and calculations on larger systems have not yet been performed.

In this paper, we consider the MLCCSD and CCSD-in-HF nitrogen K-edge spectra of adenine, adenosine, adenosine triphosphate (ATP), and an ATP-water system. 
\rev{As} core excitation processes are spatially localized, \rev{the orbital space can be partitioned} using occupied Cholesky orbitals and PAOs.
\revnew{For valence excitations, which are generally more delocalized in character, orbital selection can be more challenging and active spaces determined from CNTOs are often preferable, except when the excitation of interest is localized in some known region of the system, e.g. in solvent-solute systems. The CCSD-in-HF approach relies on the selection of an active region and to treat delocalized valence excitations, a reduced space approach like LoFEx/CorNFLEx is more appropriate.}

With the calculations presented in this paper, we outline a procedure to obtain accurate NEXAFS spectra for larger molecular systems, liquids, or solutions.
First, a model system is used to determine the basis set and to ensure that the active space of the MLCCSD and CCSD-in-HF calculations is suitable for accurate treatment of the core excitations. Here, we use adenine and adenosine for this purpose. 
Afterwards, the MLCCSD and CCSD-in-HF calculations are performed on the full system, i.e., ATP and the ATP-water system.
The systems were selected because experimental spectra are available\citep{plekan2008theoretical,kelly2010communication} and because ATP (C$_{10}$H$_{16}$N$_5$O$_{13}$P$_3$) is large enough that the full CCSD NEXAFS spectra is computationally expensive to generate.
Another theoretical NEXAFS study on adenine and ATP\rev{, both} in vacuum and in aqueous solution\rev{,} has been performed at the DFT level of theory, using polarizable density embedding to describe the solvent.\citep{hrsak2018one}  \rev{In that study,}  a series of representative geometries for ATP solvated in water \rev{were considered; this} is likely necessary
in order accurately describe the NEXAFS spectra of a solute. 
Our mission in the present study is not an accurate description of the experiment, but rather to establish the performance of \rev{the} EOM-MLCCSD and EOM-CCSD-in-HF implementations and their usefulness for modeling the NEXAFS spectra of complex systems. 

\section{Theory}
The coupled cluster wave function is given by
\begin{align}
    \ket{\mrm{CC}} = \exp(X)\ket{\mrm{HF}}, \quad X = \sum_{\mu} x_\mu \tau_\mu,
\end{align}
where $X$ is the cluster operator,
$\ket{\mrm{HF}}$ is the Hartree-Fock reference, 
$x_\mu$ are the cluster amplitudes, 
and $\tau_\mu$ are excitation operators.
The standard models within the coupled cluster hierarchy are obtained by restricting $X$ to include excitation operators up to a certain order.
The cluster amplitudes are determined through the projected coupled cluster equations,
\begin{align}
   \Omega_{\mu} = \Tbraket{\mu}{\bar{H}}{\mrm{HF}} = 0,\label{eq:omega}\\
\end{align}
and the energy is obtained from
\begin{align}
    E_{\mrm{CC}} = \Tbraket{HF}{\bar{H}}{\mrm{HF}},\label{eq:energy}
\end{align}
where $\bar{H} = \exp(-T)H\exp(T)$ and 
\begin{align}
    H = \sum_{pq}h_{pq}E_{pq} + \frac{1}{2}\sum_{pgrs}g_{pqrs}\Big(E_{pq}E_{rs} - E_{ps}\delta_{qr}\Big) + h_{\text{nuc}}
\end{align}
is the non-relativistic electronic Hamiltonian operator in terms of the singlet operators $E_{pq}$.\citep{Helgaker:2014aa} The $g_{pqrs} = (pq|rs)$ are the electron repulsion integrals in the Mulliken notation.

\subsection{Equation-of-motion coupled cluster theory}
In the equation-of-motion (EOM) coupled cluster framework,\citep{sekino1984linear,geertsen1989equation,bartlett2012coupled,krylov2008equation} a general state is expressed as
\begin{align}
\begin{split}
    \ket{k}&=
    e^{T}(\ket{\mrm{HF}}R^k_0 + \sum_{\mu>0} \ket{\mu}R^k_{\mu})\\
    &=\sum_{\mu\geq0}e^{T}R^k_{\mu}\ket{\mu},
    \end{split}
\end{align}
\rev{where $\ket{\mu} = \tau_{\mu}\ket{\mrm{HF}}$},
and $\bld{R^k}$ is obtained as the right eigenvectors of the similarity transformed Hamiltonian,
\begin{align}
    \bld{\bar{H}} =
    \begin{pmatrix}
    0 & \bld{\eta}^T \\
    0 & \bld{A} \\
    \end{pmatrix} + E_\mrm{CC}\bld{I}.
\end{align} 
Here, $\bld{A}$ is the Jacobian matrix, with elements $A_{\mu\nu}=\Tbraket{\mu}{[\bar{H},\tau_{\nu}]}{\mrm{HF}}$,
$\eta_{\nu} = \Tbraket{\mrm{HF}}{[\bar{H}, \tau_{\nu}]}{\mrm{HF}}$,
and
we have assumed that the ground state amplitudes have been determined from eq. \eqref{eq:omega}.
As $\bld{\bar{H}}$ is not Hermitian, the left eigenvectors differ from the right eigenvectors. We have the EOM coupled cluster left states
\begin{align}
   \bra{k} = \sum_{\mu\geq0}L^k_{\mu}\bra{\mu}\exp(-T)
\end{align}
and we require that the left and right eigenvectors satisfy the biorthonormalization criterion 
\begin{align}
    \Dbraket{k}{l} = \delta_{kl}.
\end{align}

The ground state solutions, $\bld L^0$ and $\bld R^0$, are given by
\begin{align}
    \bld L^0 = 
    \begin{pmatrix}
    1 \\
    \bld{l}_0\\
    \end{pmatrix},\quad
    \bld R^0 = 
    \begin{pmatrix}
    1\\
    0\\
    \end{pmatrix}, 
\end{align}
where $\bld{l_0}$ is the ground state multipliers determined from 
\begin{align}
    \bld{A}^T\bld{l}_0 = -\bld{\eta}.\label{eq:multipliers}
\end{align}{}
The excited state solutions, $\bld L^k$ and $\bld R^k$ for $k > 0$, are given by
\begin{align}
    \bld L^k = 
    \begin{pmatrix}
    0 \\
    \bld{l}_k\\
    \end{pmatrix},\quad
    \bld R^k = 
    \begin{pmatrix}
    r_0 \\
    \bld{r}_k\\
    \end{pmatrix}, 
\end{align}
where $\bld{l}_k$ and $\bld{r_k}$ are left and right eigenvectors of $\bld{A}$,
respectively,
and $r_0 = -\bld{l_0}\cdot\bld{r_k}$. The eigenvalues of $\bld{A}$ are the excitation energies, $\omega_k$.

Oscillator strengths for transitions between the ground and the $k$'th excited state are given by
\begin{align}
    f_k = \frac{2}{3}\omega_k\Tbraket{0}{\bld\mu}{k}\cdot\Tbraket{k}{\bld\mu}{0},\label{eq:os1}
\end{align}
where $\mu^{\alpha} = \sum_{pq}\mu^{\alpha}_{pq}E_{pq}$ is the $\alpha$ component of dipole operator.\citep{stanton1993equation,koch1994calculation}

\subsection{Multilevel CCSD}
In multilevel coupled cluster theory, we restrict the higher order excitations of the cluster operator to an active orbital space.
In the two-level CCS/CCSD approach, the cluster operator assumes the form
\begin{align}
    X^{\mrm{MLCCSD}} = X_1 +  T_2.
\end{align}
The single excitation operator,
\begin{align}
    X_1 = \sum_{\mu_1}x_{\mu_1}\tau_{\mu_1} = \sum_{AI}x^{A}_{I}E_{AI},
\end{align}
includes single excitations in the entire orbital space, that is,
the summation indices $A$ and $I$ label general (active and inactive) virtual and occupied orbitals, respectively.
The double excitation operator $T_2$ is restricted to the active orbital space,
\begin{align}
    T_2 = \sum_{\mu_2}t_{\mu_2}\tau_{\mu_2} 
    =  \frac{1}{2}\sum_{aibj}t^{ab}_{ij}E_{ai}E_{bj},\label{eq:t2}
\end{align}
where the summation indices $a, b$ and $i, j$ label active virtual and occupied orbitals, respectively.

The MLCCSD ground state equations for the two-level CCS/CCSD model are
\begin{align}
    \Omega_{\mu_1} = &\Tbraket{\mu_1}{\hat{H} + [\hat{H}, T_2]}{\mrm{HF}} = 0\label{eq:mlccsd_1}\\
    \Omega_{\mu_2} = &\Tbraket{\mu_2}{\hat{H} + [\hat{H}, T_2] + \frac{1}{2} [[\hat{H}, T_2], T_2]}{\mrm{HF}} = 0,\label{eq:mlcc2_2}
\end{align}
where $\hat{H}$ is the $X_1$-transformed Hamiltonian and the doubles projection space is associated with $T_2$. These equations are equivalent to the standard CCSD ground state equations, except for the restriction of the $T_2$ operator to the active space.

Properties of this MLCCSD model can be obtained within the EOM framework.
The excited states ($\ket{k}$, $\bra{k}$) are constructed by solving the eigenvalue equations of the MLCCSD (CCS/CCSD) Jacobian matrix,
\begin{align}
    \bld{A}^{\text{MLCCSD}}\bld{r}_k = \omega_k\bld{r}_k\\
    (\bld{A}^{\text{MLCCSD}})^T\bld{l}_k = \omega_k\bld{l}_k,
\end{align}
where
\begin{align}
  &\bld A^{\mrm{MLCCSD}} =
  &\begin{pmatrix}
    \Tbraket{\mu_1}{[\hat{H},\tau_{\nu_1}]+ [[\hat{H}, \tau_{\nu_1}], T_2]}{R} &
    \Tbraket{\mu_1}{[\hat{H}, \tau_{\nu_2^{\mrm{T}}}]}{R} \\
    \Tbraket{\mu_2}{[\hat{H},\tau_{\nu_1}]+ [[\hat{H}, \tau_{\nu_1}], T_2]}{R} &
    \Tbraket{\mu_2}{[\hat{H},\tau_{\nu_2}]+ [[\hat{H}, \tau_{\nu_2}], T_2]}{R}
  \end{pmatrix}.
\end{align}
\rev{Note that $\bld{A}^{\mrm{MLCCSD}}$ assumes the same form as $\bld{A}^{CCSD}$, except for restriction of the operator $T_2$ and the corresponding projection space.}
Core excited states can be obtained using the core-valence separation (CVS) approach of Coriani and Koch.\citep{coriani2015communication,coriani2016erratum}
In this approach, the non-zero elements of the excitation vectors have at least one occupied index belonging to the excited core orbital. It can be implemented as a projection \citep{coriani2015communication,coriani2016erratum} or by implementing the linear transformation by the CVS Jacobian matrix directly.\citep{vidal2019new,alex2020new}
Once the left and right states are constructed and the multipliers are determined from eq \eqref{eq:multipliers}, then the MLCCSD oscillator strengths can be calculated according to eq \eqref{eq:os1}.

\subsection{Partitioning the orbital space}
The first step of any multilevel coupled cluster calculation is to partition the molecular orbitals into the active and inactive orbital sets. 
The canonical Hartree-Fock orbitals are not suitable to determine the active space. 
If the property of interest is spatially localized in the molecular system, such as core excitations or excitations in a target molecule in a solvent, localized orbitals can be used.

For the occupied space, there are many widely used iterative localization procedures, such as the Boys,\citep{boys1960construction} Pipek-Mezey,\citep{pipek1989fast} and Edmiston-Ruedenberg\citep{edmiston1963localized} procedures. In this work, we use the semilocalized Cholesky orbitals described in Refs.~\citenum{aquilante2006fast} and \citenum{sanchez2010cholesky}, which can be obtained in a non-iterative procedure. A set of active atoms are selected and the idempotent Hartree-Fock density,
\begin{align}
    D_{\alpha\beta} = \sum_{I}C_{\alpha I}C_{\beta I},
\end{align}
is Cholesky decomposed in a specialized procedure where the pivoting elements are restricted to correspond to AOs on the active atoms.
The decomposition procedure ends when all ``active'' diagonals fall below a given threshold. After the decomposition, the Cholesky factors are the orbital coefficients of the active occupied orbitals, $\bld{C}^{\text{a}}$:
\begin{align}
    \bld{D} = \bld{C}^{\text{a}}(\bld{C}^{\text{a}})^T + \bld{D}^{\text{e}}.
\end{align}
The density of the inactive space, $\bld{D}^{\text{e}}$, can be fully Cholesky decomposed to yield the inactive occupied orbitals. 

The iterative localization procedures which are extensively used for the occupied space can also be applied to the virtual space. However, convergence for the virtual space is more challenging and the use of sophisticated level-shift and trust-radius solvers are often necessary.\citep{hoyvik2016characterization}
The projected atomic orbitals (PAOs), is an alternative to such iterative localization procedures for the virtual space.
To construct PAOs in an active region of the molecular system, the occupied orbitals are projected out of the AOs centered on the active atoms.
The orbital coefficient matrix for the active virtual PAOs is 
\begin{align}
\bld C^{\text{PAO}} = \bld I - \bld{DS'},
\end{align}
where $\bld S'$ is rectangular and contains the columns of the AO overlap matrix that correspond to AOs centered on the active atoms. These orbitals are non-orthogonal and linearly dependent. 
The Löwdin canonical orthonormalization procedure,\citep{lowdin1970nonorthogonality} can be used to obtain a set of orthonormal active virtual orbitals. 
Linear dependence in the full set of AOs should be removed before the PAO construction.
The inactive virtual orbitals are obtained in a similar way. 
The active virtual orbitals, as well as the occupied orbitals, are projected out of the full set of AOs. The resulting orbitals are orthonormalized. 

After the orbitals have been partitioned, 
we block diagonalize the occupied-occupied and virtual-virtual Fock matrices such that the active-active and inactive-inactive blocks become diagonal. This is achieved by rotating among the active orbitals and among the inactive orbitals, separately.
This semicanonical basis is used throughout the MLCCSD calculation, as this significantly improves convergence.

\subsection{Reduced space CCSD}
In MLCCSD, the double excitations included in the cluster operator are restricted to an active space. In the reduced space CCSD-in-HF approach, the single excitations are also restricted to the active space. The active orbitals are determined from occupied Cholesky orbitals and PAOs. 
The inactive orbitals contribute through the Fock matrix, 
\begin{align}
   \begin{split}
   F_{pq} &= h_{pq} + \sum_{i}\left(2g_{pqii} - g_{piiq}\right) +
   \sum_{\mathcal{I}} \left(2g_{pq\mathcal{I}\mathcal{I}} - g_{p\mathcal{I}\mathcal{I}q}\right)\\
   &= h_{pq} + \sum_{i}\left(2g_{pqii} - g_{piiq}\right) + F_{pq}^{\mrm{e}},
   \end{split}\label{eq:effective_fock}
\end{align}
where $p$ and $q$ are general active space indices, $i$ denotes an active occupied orbital, the index $\mathcal{I}$ denotes an inactive occupied orbital.
The CCSD-in-HF calculation is performed as a standard CCSD calculation, but with a truncated MO basis consisting of the active orbitals ($N_\mrm{MO}<N_\mrm{AO}$) and with the effective Fock matrix of eq \eqref{eq:effective_fock}.

\section{Results and discussion}
\begin{figure}
    \centering
    \includegraphics[width=\linewidth]{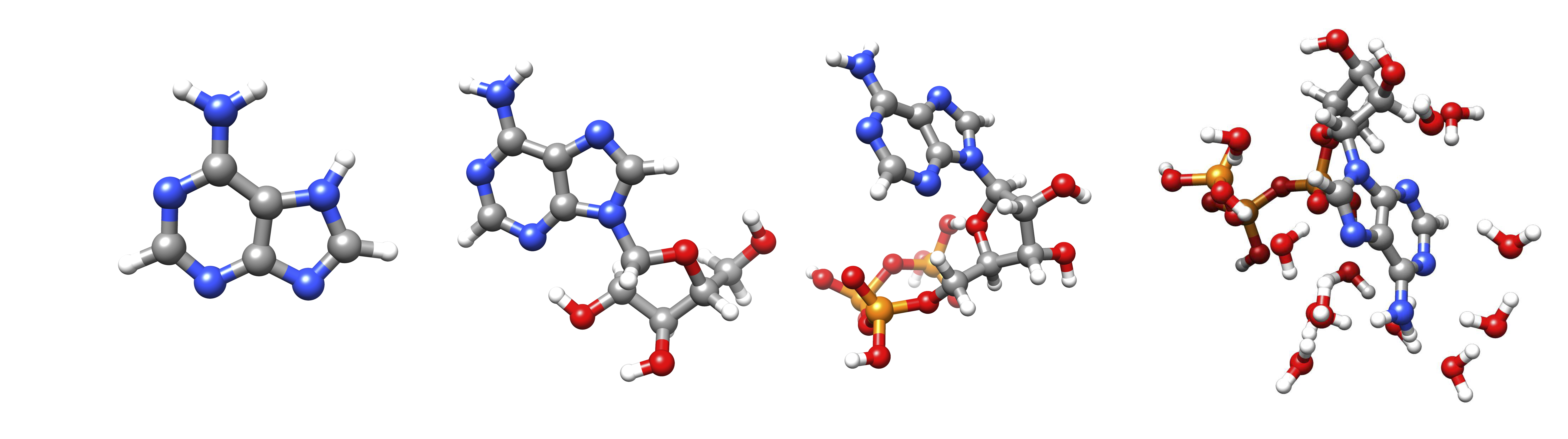}
    \caption{Adenine, adenosine, adenosine triphosphate (ATP), and ATP with twelve water molecules.}
    \label{fig:system}
\end{figure}
As a test study for the EOM-MLCCSD and EOM-CCSD-in-HF implementations, we consider the nitrogen core excitations of adenine, adenosine, and adenosine triphosphate (ATP) \textit{in vacuo} and ATP with twelve water molecules, see Figure \ref{fig:system}. 
These systems are chosen because of their biological importance and the availability of experimental studies.\citep{plekan2008theoretical, kelly2010communication} 
In particular, experimental NEXAFS spectra at the nitrogen and carbon edge of adenosine triphosphate in aqueous solution has been reported.\citep{kelly2010communication}
Our goal in this paper is not to perform an accurate application study, but rather to demonstrate the performance, in terms of accuracy and cost, of the MLCCSD and CCSD-in-HF methods. 
This has dictated our choice of basis sets and the number of computed states.
Furthermore, in order to properly describe the effects of solvents, one should sample the spectra at several representative geometries, e.g. obtained from a molecular dynamics simulation. 
Bulk solvent should also be included in the system, for instance by using the QM/MM framework\citep{warshel1972,levitt1975,warshel1976}, polarizable continuum model,\citep{tomasi2005quantum,mennucci2012polarizable} or by treating all water molecules at the Hartree-Fock level of theory.
In general, triple excitations (CC3\citep{koch1997cc3}) are needed to obtain quantitative, unshifted NEXAFS spectra; this is demonstrated for adenine.
However, shifted CCSD spectra can be useful for qualitative interpretation of experiments.

All geometries, except the ATP-water \rev{and methylamine-water geometries}, are obtained at the B3LYP/aug-cc-pVDZ level using the NWChem\citep{valiev2010nwchem} software. The ATP-water \rev{and methylamine-water geometries were} built using the Avogadro software package.\citep{hanwell2012avogadro} All geometries are available from Ref.~\citenum{geometries}. 
Visualization of the molecular systems is done using the Chimera software package.\citep{pettersen2004ucsf}

The EOM-MLCCSD oscillator strengths were implemented in a development branch of the eT program\citep{eTpaper} 
and all calculations are performed with eT.
\rev{The following thresholds have been used:} For the ground state, we used threshold of $10^{-6}$ on $|\bld\Omega|$ and on the residual of the multiplier equations. For the excited states, a threshold of $10^{-4}$ was used for the residual and $10^{-6}$ on the change in the excitation energies.
The electron repulsion integrals are Cholesky decomposed and the decomposition threshold is \rev{ $10^{-8}$ for adenine and $10^{-6}$ for the remaining calculations. Generally, a decomposition threshold of $10^{-3}$ is sufficient for accurate excitation energies.}
All timings were performed on two Intel Xeon Gold 6138 processors, using 40 threads\rev{, and ${360}$ GB of memory was available} in all calculations.
In the EOM-MLCCSD and EOM-CCSD-HF calculations, we use Cholesky-PAOs to partition the orbital space and the ``adenine part'' of adenosine and ATP is considered active.

\section{Adenine}
\begin{figure}
    \centering
    \includegraphics[width=0.9\linewidth]{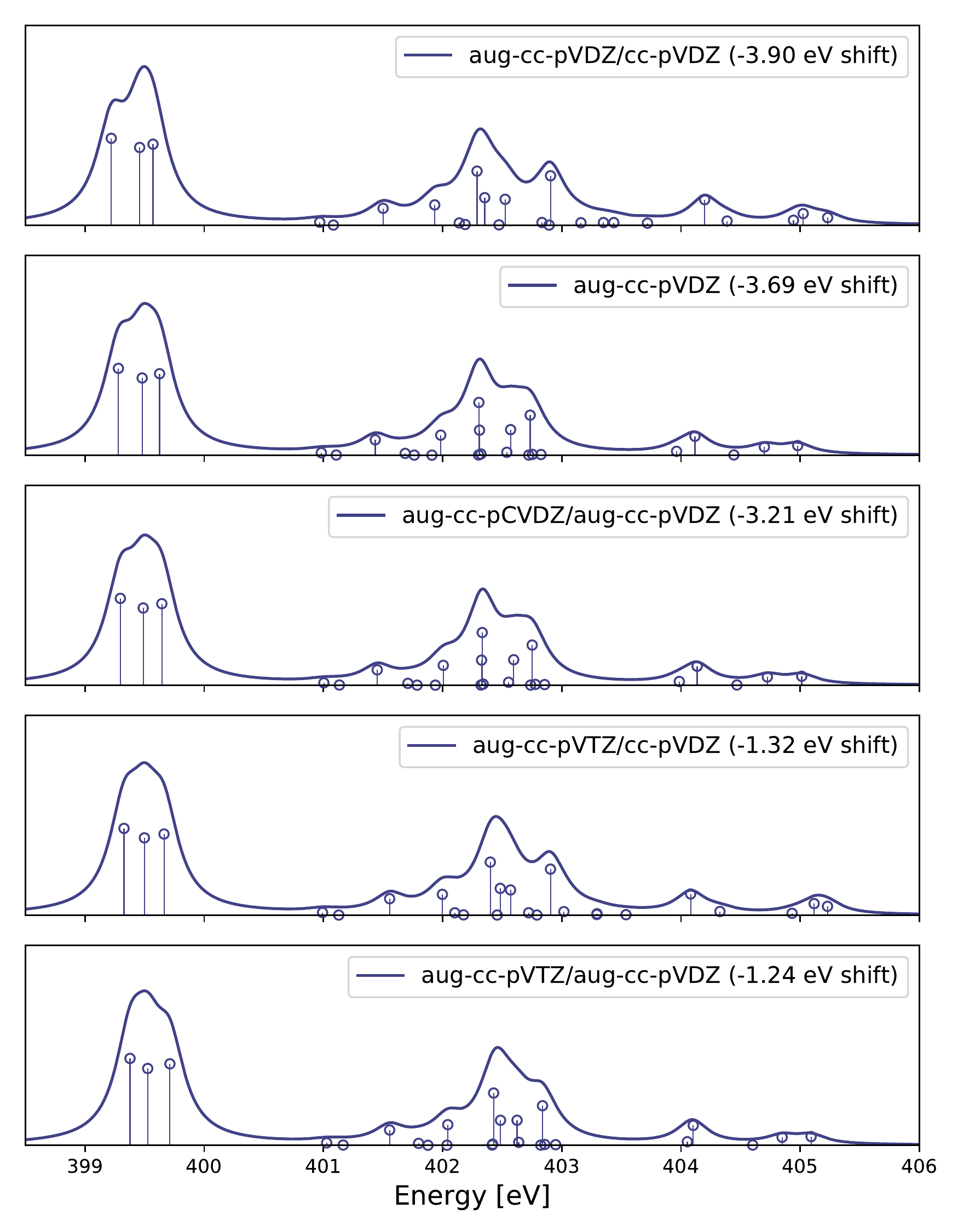}
    \caption{Adenine NEXAFS nitrogen edge calculated at the CCSD level of theory with combinations of the Dunning correlation consistent basis sets. When a larger basis set is used on the nitrogen atoms, we use the notation: \textit{basis-on-nitrogen/basis-on-other-atoms}.  Lorentzian broadening with \rev{$\SI{0.3}{\eV}$} FWHM has been applied. The first peak has been shifted to the experimental value $\SI{399.5}{\eV}$, as reported in Ref.~\citenum{plekan2008theoretical}. Six roots was calculated for each nitrogen atom.}
    \label{fig:Adenine_CCSD}
\end{figure}

\begin{figure}
    \centering
    \includegraphics[width=0.7\linewidth]{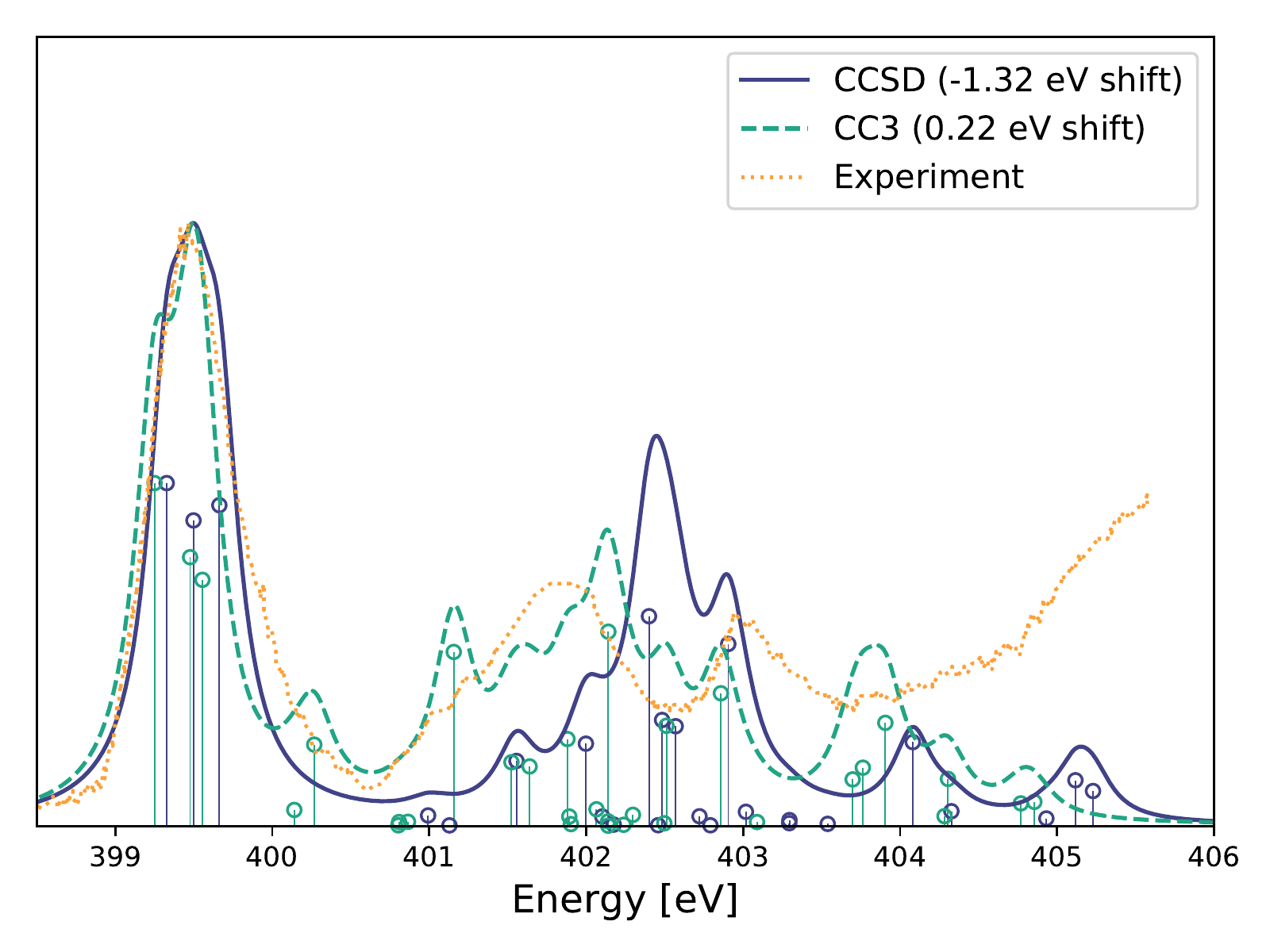}
    \caption{Adenine NEXAFS nitrogen edge calculated at the CCSD and CC3 level of theory with the aug-cc-pVTZ/cc-pVDZ basis set. The larger basis set is used on the nitrogen atoms. The first peak has been shifted to the experimental value $\SI{399.5}{\eV}$, as reported in Ref.~\citenum{plekan2008theoretical}. Six roots have been calculated for each nitrogen atom.
     Lorentzian broadening with \rev{$\SI{0.3}{\eV}$} FWHM has been applied. 
    Experimental data collected from Ref.~\citenum{plekan2008theoretical} using WebPlotDigitizer.\citep{webplotdigitizer}}
    \label{fig:Adenine_CCSD_CC3}
\end{figure}
In order to select the basis set for the larger systems, we start by considering the EOM-CCSD nitrogen K-edge spectrum of adenine for different combinations of Dunning correlation consistent basis sets.\citep{dunning1989gaussian,kendall1992electron}
The results are given in Figure \ref{fig:Adenine_CCSD}.
Generally, we use a larger basis set on the nitrogen atoms than on the carbon and hydrogen atoms.
From the spectra with aug-cc-pVDZ and aug-cc-pCVDZ/aug-cc-pVDZ, 
we observe that additional core functions on the nitrogen atoms shifts the spectrum, but that the overall shape is unchanged.
From the two lower panels, 
we see that the shift, with respect to the experimental value of $\SI{399.5}{\eV}$ for the first peak,\citep{plekan2008theoretical} is significantly reduced by using triple-zeta rather than double-zeta on the nitrogen atoms.
Although the finer details might differ, 
the main features of the spectrum are the same for all basis set combinations. 
As a compromise between accuracy and cost, we use the aug-cc-pVTZ/cc-pVDZ basis sets for the remaining calculations.

In Figure \ref{fig:Adenine_CCSD_CC3}, the \rev{EOM-}CCSD/aug-cc-pVTZ/cc-pVDZ and \rev{EOM-}CC3/aug-cc-pVTZ/cc-pVDZ\citep{alex2020new} are compared to the experimental spectrum from Ref.~\citenum{plekan2008theoretical}.
The inclusion of triple excitations significantly improves the computed spectrum for the second feature at approximately $\SI{402}{\eV}$.
Furthermore, more than six roots per nitrogen atom are necessary in order to describe the spectrum from $\SI{403}{\eV}$. To accurately describe Rydberg states, additional diffuse basis functions are likely needed. 

\section{Adenosine}
\begin{figure}
    \centering
    \includegraphics[width=0.7\textwidth]{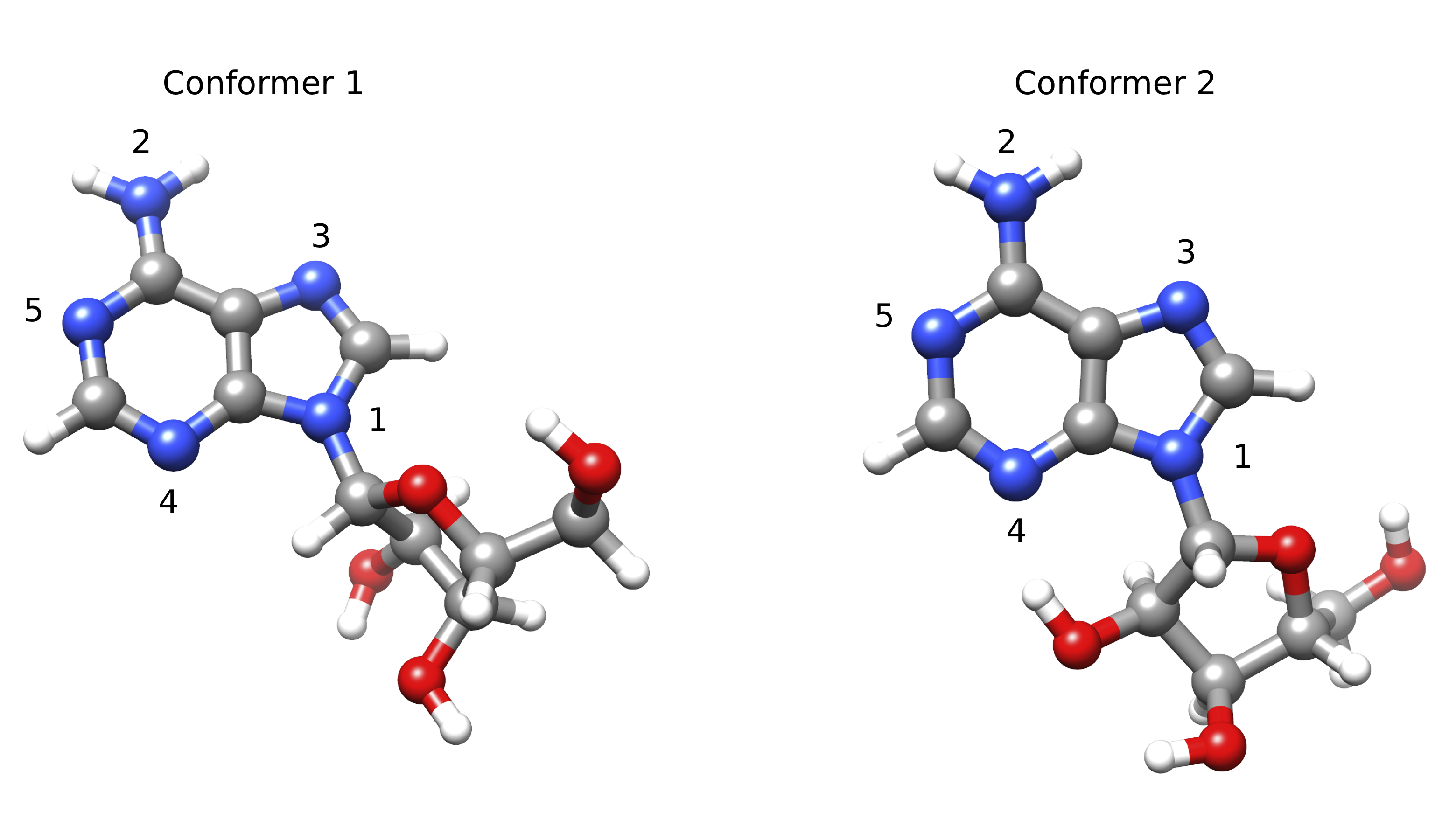}
    \caption{Adenosine conformers 1 and 2 with labels corresponding to Tables \ref{tab:Adenosine_es_energies} and \ref{tab:Adenosine_es_energies_ccsd_in_hf}.
    \label{fig:Adenosine_numbers}
    }
\end{figure}

\begin{figure}
    \centering
    \includegraphics[width=0.8\textwidth]{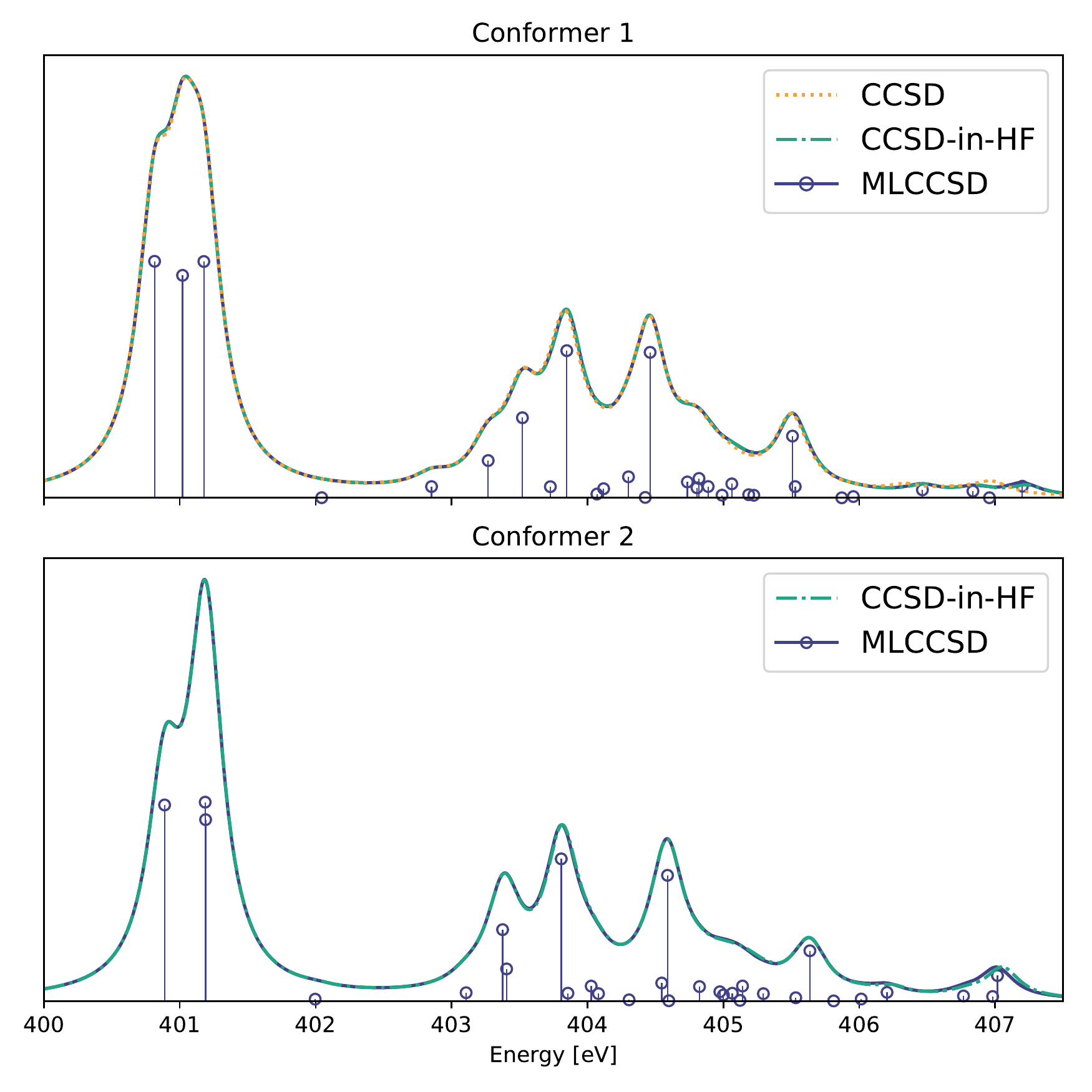}
    \caption{Nitrogen K-edge NEXAFS spectra of two conformers of Adenosine calculated at the EOM-MLCCSD/aug-cc-pVTZ/cc-pVDZ and EOM-CCSD-in-HF/aug-cc-pVTZ/cc-pVDZ levels of theory. Six roots have been calculated for each nitrogen atom.
     Lorentzian broadening with \rev{$\SI{0.3}{\eV}$} FWHM has been applied.}
    \label{fig:Adenosine_MLCCSD}
\end{figure}

To illustrate the performance of EOM-MLCCSD and EOM-CCSD-in-HF compared to EOM-CCSD for core excitations and oscillator strengths, we consider the nitrogen edge NEXAFS spectrum of adenosine (conformer 1 in Figure \ref{fig:Adenosine_numbers}) calculated using the aug-cc-pVTZ/cc-pVDZ basis. The results are given in Figure \ref{fig:Adenosine_MLCCSD}, Table \ref{tab:Adenosine_es_energies}, and Table \ref{tab:Adenosine_es_energies_ccsd_in_hf}.
In the MLCCSD and CCSD-in-HF calculations for conformer 1, there are 40 active occupied orbitals and 301 active virtual orbitals, and there are 30 inactive occupied orbitals and 120 inactive virtual orbitals. 
For nitrogen atoms 2--5 (see Table \ref{tab:Adenosine_es_energies}), the calculated MLCCSD and CCSD-in-HF excitation energies for adenosine has an error with respect to CCSD of less than $\SI{0.1}{eV}$. This is well within the expected error of CCSD for K-edge core excitations.
The errors are generally larger for nitrogen atom 1 (N1). \rev{
This is because} not all nearest neighbours of N1 are defined as active. 
The errors for N1 can be reduced by including the neighbouring carbon on the ribose into the set of active atoms, as seen from Table \ref{tab:Adenosine_es_energies}.
As seen from Figure \ref{fig:Adenosine_MLCCSD} and Tables  \ref{tab:Adenosine_es_energies} and \ref{tab:Adenosine_es_energies_ccsd_in_hf}, \rev{the errors of MLCCSD and CCSD-in-HF, compared to full CCSD, are small: t}he spectra coincide for all but the low intensity peaks at around $\SI{407}{\eV}$. The MLCCSD and CCSD-in-HF spectra are almost indistinguishable, with small differences observed for the high energy excitations only.
\begin{table}
\scriptsize
    \centering
    \begin{threeparttable}
    \begin{tabular}{c c c c c c c}
    \toprule
    Nitrogen & $\omega_1\;(\Delta\omega_1)$ & $\omega_2\;(\Delta\omega_2)$ & $\omega_3\;(\Delta\omega_3)$ & $\omega_4\;(\Delta\omega_4)$ & $\omega_5\;(\Delta\omega_5)$ & $\omega_6\;(\Delta\omega_6)$  \\
    \midrule
    N1 & 403.848 (0.011) & 405.529 (0.041) & 406.465 (0.130) & 406.836 (0.207) & 406.958 (0.120) & 407.200 (0.219) \\
    N2 & 403.269 (0.004) & 403.521 (0.007) & 404.461 (0.004) & 405.509 (0.006) & 405.871 (0.012) & 405.956 (0.009)\\
    N3 & 400.815 (0.004) & 404.119 (0.031) & 404.302 (0.013) & 404.821 (0.020) & 404.990 (0.068)& 405.186 (0.053)\\
    N4  & 401.021 ($<$0.001) & 402.853 (0.001) & 404.069 (0.026) & 404.806 (0.022) & 405.062 (0.046) & 405.224 (0.028) \\
    N5 & 401.177 ($<$0.001) & 402.044 ($<$0.001) & 403.727 (0.004) & 404.425 (0.006) & 404.735 (0.013) & 404.888 (0.007) \\
    \midrule 
    N1\tnote{$\dagger$} & 403.845 (0.008) &  405.503 (0.015)  & 406.407 (0.072) & 406.755 (0.126) & 406.915 (0.077) & 407.104 (0.123)\\
    \bottomrule
    \end{tabular}
    \caption{MLCCSD/aug-cc-pVTZ/cc-pVDZ core excitations out of the nitrogen 1s orbitals for adenosine (conformer 1). Excitation energies, $\omega_i$, and errors with respect to CCSD, $\Delta\omega_i$, are given in eV.}
    \label{tab:Adenosine_es_energies}
    \begin{tablenotes}
      \item[$\dagger$] {Nearest neighbouring carbon on ribose included in active space.}
    \end{tablenotes}
    \end{threeparttable}
\end{table}{}

\begin{table}
\scriptsize
    \centering
    \begin{tabular}{c c c c c c c}
    \toprule
    Nitrogen & $\omega_1\;(\Delta\omega_1)$ & $\omega_2\;(\Delta\omega_2)$ & $\omega_3\;(\Delta\omega_3)$ & $\omega_4\;(\Delta\omega_4)$ & $\omega_5\;(\Delta\omega_5)$ & $\omega_6\;(\Delta\omega_6)$  \\
    \midrule
    N1 & 403.850 (0.013) & 405.526 (0.038) & 406.474 (0.140) & 406.854 (0.224) & 406.961 (0.123) & 407.232 (0.251) \\
    N2 & 403.269 (0.004) & 403.522 (0.008) & 404.461 (0.004) & 405.508 (0.004) & 405.874 (0.016) & 405.957 (0.010) \\
    N3 & 400.815 (0.004) & 404.119 (0.031) & 404.300 (0.011) & 404.818 (0.018) & 404.998 (0.076) & 405.187 (0.054) \\
    N4 & 401.021 ($<$0.001) & 402.855 (0.002) & 404.074 (0.032) & 404.807 (0.023) & 405.068 (0.052) & 405.226 (0.030)\\
    N5 & 401.178 (0.001) & 402.045 (0.001) & 403.726 (0.003) & 404.426 (0.007) & 404.736 (0.014) & 404.888 (0.008)\\
    \bottomrule
    \end{tabular}
    \caption{CCSD-in-HF/aug-cc-pVTZ/cc-pVDZ core excitations out of the nitrogen 1s orbitals for adenosine (conformer 1). Excitation energies, $\omega_i$, and errors with respect to CCSD, $\Delta\omega_i$, are given in eV.}
    \label{tab:Adenosine_es_energies_ccsd_in_hf}
\end{table}{}

Wall times are given in Table \ref{tab:Adenosine_wall_times}. We report average timings from the calculations of the nitrogen atoms for the cluster amplitudes ($t^{\text{gs}}$) and the average time to transform by the Jacobian matrix, $\bld A$, and $\bld A^T$ ($t^{{\bld A}}$, and $t^{{\bld A^T}}$). Although the active space is quite large in these calculations, the computational savings are significant with approximately a factor five for $t^{\text{gs}}$, $t^{{\bld A}}$, and $t^{{\bld A^T}}$, compared to full CCSD.
Specialized active spaces for each of the five MLCCSD and CCSD-in-HF calculations can be used where we include only the neighbours of the nitrogen atom being excited. This can significantly reduce the cost, but will probably lead to increased errors with respect to the full CCSD calculation. 

The time to converge the excited states varies greatly because the number of iterations required to reach convergence varies for the different nitrogen atoms. 
To give a perspective on the computational savings achieved in these MLCCSD calculations we compare the calculation time of the cheapest CCSD calculation (N2) with the most expensive MLCCSD calculation (N4); the cheapest CCSD calculation used 9 days and 16 hours, whereas the most expensive MLCCSD calculation used 1 day and 8 hours. There are two contributing factors to the savings with MLCCSD: the savings due to the reduced cost of constructing the $\bld{\Omega}$-vector and performing the linear transformation by $\bld{A}$ and  $\bld{A}^T$, and the reduction of IO in the MLCCSD calculations. The IO is reduced because the reduced space of the Davidson procedure can be \rev{stored} in memory for MLCCSD, whereas this is not possible for CCSD with 360 GB of memory available. 
\begin{table}
    \centering
    \begin{tabular}{l c c c}
    \toprule
    &$t^{\text{gs}}~[\si{\minute}]$ & $t^{{\bld A}}~[\si{\minute}]$ &
    $t^{{\bld A^T}}~[\si{\minute}]$  \\
    \midrule
    CCSD-in-HF & 19  & 2 & 3  \\
    MLCCSD     & 22  & 2 & 3  \\ 
    CCSD       & 128 & 11 & 14 \\
    \bottomrule
    \end{tabular}
    \caption{Average wall times to solve for the ground state equations, $t^{\text{gs}}$ and average time for transformations by $\bld{A}^T$ ($t^{{\bld A^T}}$) and $\bld{A}$ ($t^{{\bld A}}$) of adenosine (conformer 1).}
    \label{tab:Adenosine_wall_times}
\end{table}

In Figure \ref{fig:Adenosine_MLCCSD}, we also present the EOM-MLCCSD/aug-cc-pVTZ/cc-pVDZ and EOM-CCSD-in-HF/aug-cc-pVTZ/cc-pVDZ  nitrogen K-edge spectra for conformer 2 of adenosine.
For conformer 2, there are 42 active occupied orbitals and 303 active virtual orbitals, and there are 28 inactive occupied orbitals and 118 inactive virtual orbitals. We observe a change in the spectrum, compared to conformer 1, resulting from the change in environment of the nitrogen atoms. Again, the MLCCSD and CCSD-in-HF models give very similar results, as seen for conformer 1.
\section{ATP and ATP-water}
\begin{figure}
    \centering
    \includegraphics[width=0.7\textwidth]{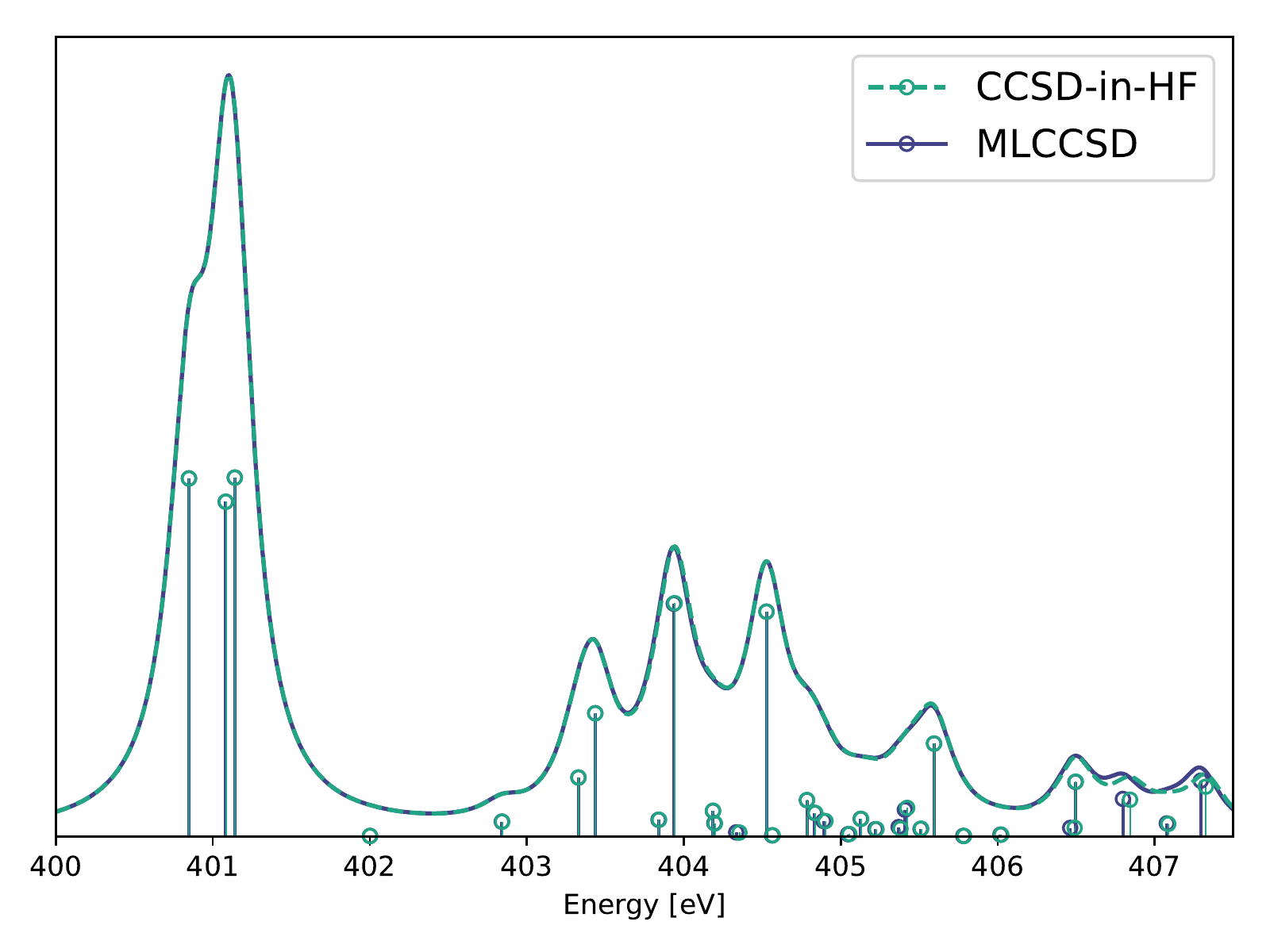}
    \caption{Nitrogen K-edge NEXAFS spectra calculated at the MLCCSD/aug-cc-pVTZ/cc-pVDZ and CCSD-in-HF/aug-cc-pVTZ/cc-pVDZ level for ATP. Six roots have been calculated for each nitrogen atom.
     Lorentzian broadening with \rev{$\SI{0.3}{\eV}$} FWHM has been applied. }
    \label{fig:ATP_MLCCSD}
\end{figure}
\begin{figure}
    \centering
    \includegraphics[width=0.7\textwidth]{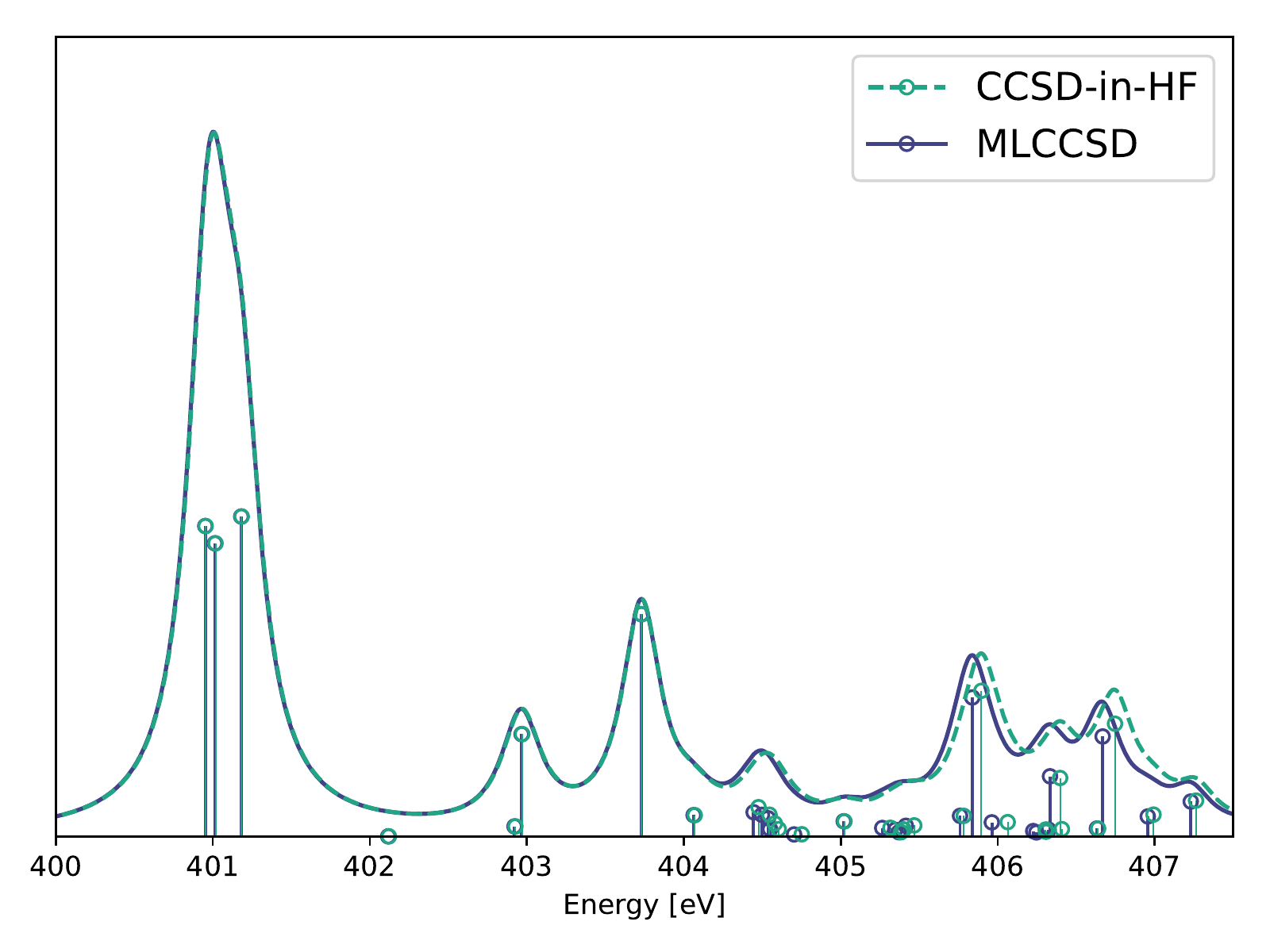}
    \caption{Nitrogen K-edge NEXAFS spectra calculated at the MLCCSD/aug-cc-pVTZ/cc-pVDZ and CCSD-in-HF/aug-cc-pVTZ/cc-pVDZ  level for ATP and 12 water molecules. Six roots have been calculated for each nitrogen atom.
     Lorentzian broadening with \rev{$\SI{0.3}{\eV}$} FWHM has been applied. }
    \label{fig:ATP_water_MLCCSD}
\end{figure}
The EOM-MLCCSD and EOM-CCSD-in-HF methods can be used to treat systems for which full EOM-CCSD is too expensive. The EOM-MLCCSD/aug-cc-pVTZ/cc-pVDZ and EOM-CCSD-in-HF/aug-cc-pVTZ/cc-pVDZ nitrogen K-edge spectrum of ATP is given Figure \ref{fig:ATP_MLCCSD}. There are 44 active occupied orbitals and 303 active virtual orbitals, and 86 inactive occupied orbitals and 253 inactive virtual orbitals. The most expensive of the five MLCCSD calculations for this system \rev{completed in} 2 days and 23 hours.
As is the case for adenosine, the MLCCSD and CCSD-in-HF spectra coincide for all but the high energy excitations around $\SI{407}{\eV}$.

\begin{table}
    \centering
    \begin{tabular}{l c c}
    \toprule
    & $t^{{\bld A}}~[\si{\minute}]$ &
    $t^{{\bld A^T}}~[\si{\minute}]$  \\
    \midrule
    CCSD-in-HF  & 5 & 7 \\
    MLCCSD      & 33 & 37 \\ 
    \bottomrule
    \end{tabular}
    \caption{MLCCSD and CCSD-in-HF calculations on the ATP-water system using the aug-cc-pVTZ/cc-pVDZ basis. Average wall times to transformations by $\bld{A}^T$ ($t^{{\bld A^T}}$) and $\bld{A}$ ($t^{{\bld A}}$).}
    \label{tab:ATRP-WATER_wall_times}
\end{table}
In Figure \ref{fig:ATP_water_MLCCSD}, we present the nitrogen K-edge spectrum of the ATP-water system (see Figure \ref{fig:system}). 
In this system, there are 58 active occupied and 310 active virtual orbitals and 132 inactive occupied and 474 inactive virtual orbitals.
The MLCCSD and CCSD-in-HF results coincide well, but some differences are observed for higher excitation energies. This is likely because MLCCSD, with CCS on the whole system, offers an improved description the more diffuse core excited states. 
Including additional Rydberg functions on the active atoms could be important for the description of these states in both methods.
The MLCCSD calculations are more expensive than the CCSD-in-HF calculations; timings for transformations by $\bld{A}$ and $\bld{A}^T$ are given in Table \ref{tab:ATRP-WATER_wall_times}.

With the calculation on ATP and the ATP-water system, we demonstrate that the MLCCSD and CCSD-in-HF approaches can be used to treat sizable molecules and to include solvent effects explicitly.
To fully capture the effects of the solvent, the calculations must be performed on several representative geometries. Additionally, one should increase the number of water molecules included in the calculation, treating most water molecules at a lower level of theory. For such a study, the CCSD-in-HF approach is preferable, since accuracy is comparable to MLCCSD, but the cost is significantly lower.

\begin{figure}
    \centering
    \includegraphics[width=0.7\textwidth]{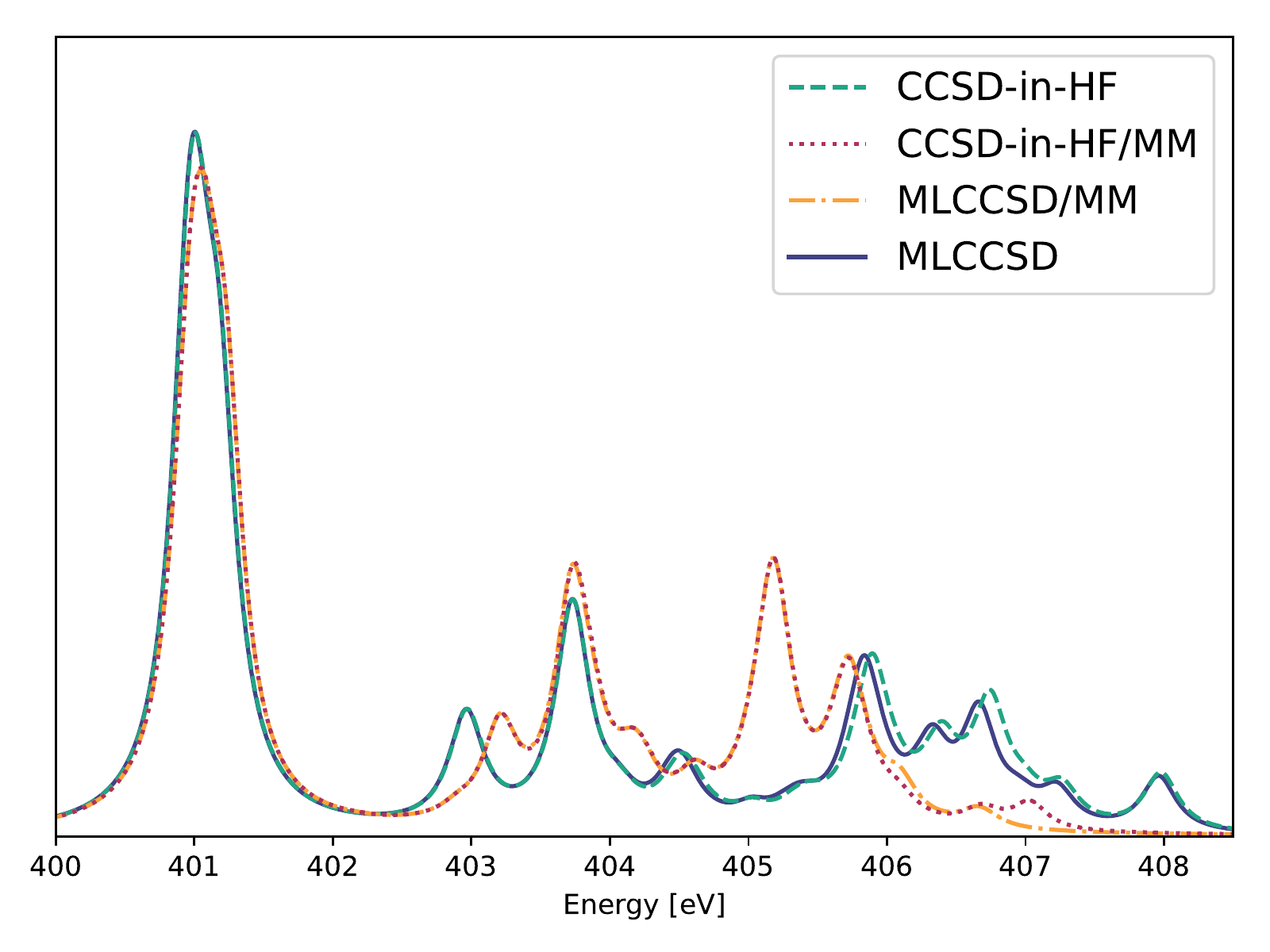}
    \caption{\rev{Nitrogen K-edge NEXAFS spectra calculated at the MLCCSD/aug-cc-pVTZ/cc-pVDZ, MLCCSD/MM/aug-cc-pVTZ/cc-pVDZ, CCSD-in-HF/aug-cc-pVTZ/cc-pVDZ, and CCSD-in-HF/MM/aug-cc-pVTZ/cc-pVDZ level for ATP and 12 water molecules. Six roots have been calculated for each nitrogen atom.
    Lorentzian broadening with \rev{$\SI{0.3}{\eV}$} FWHM has been applied. }}
    \label{fig:ATP_water_MM}
\end{figure}

\rev{The MLCCSD and CCSD-in-HF models can be used in a QM/MM framework. In Figure \ref{fig:ATP_water_MM} we compare the MLCCSD and CCSD-in-HF spectra for the ATP-water system to the QM/MM approach where the QM region (ATP) is treated with either CCSD-in-HF or MLCCSD and the MM region (the 12 water molecules) is treated with electrostatic embedding.\citep{senn2009qm} There is good agreement between the pure QM and the QM/MM spectra for the peaks in the range $\SI{401}{\eV}$--$\SI{405}{\eV}$. For the remaining peaks of higher energy, the differences are significant.}

\begin{figure}
    \centering
    \includegraphics[width=0.6\textwidth]{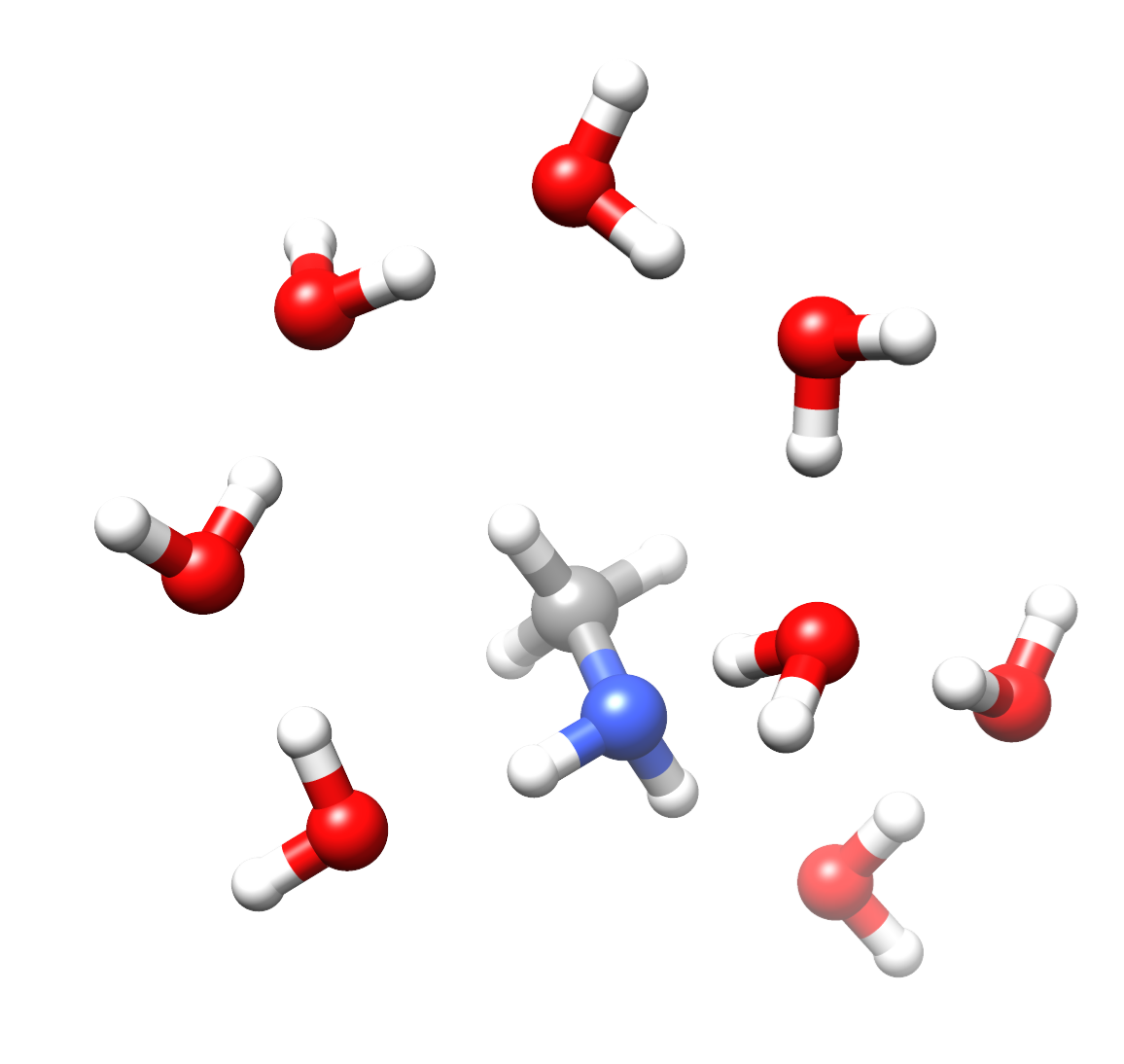}
    \caption{\rev{Methylamine and eight water molecules}}
    \label{fig:methylamine_system}
\end{figure}

\begin{figure}
    \centering
    \includegraphics[width=0.6\textwidth]{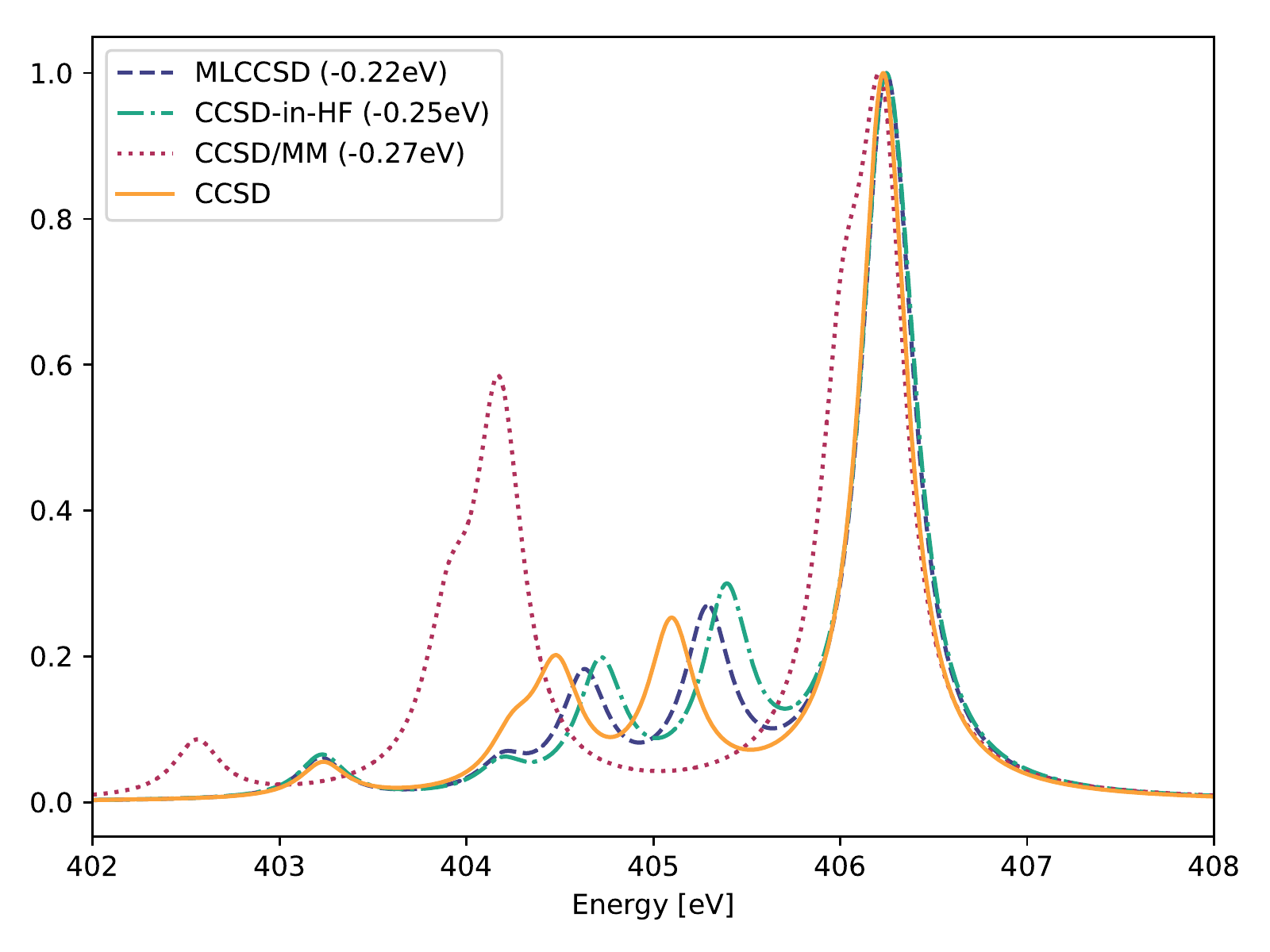}
    \caption{\rev{Nitrogen K-edge of methylamine-water system. Comparison of CCSD, MLCCSD, CCSD-in-HF and CCSD/MM using the aug-cc-pVTZ basis on the nitrogen atom and the cc-pVDZ basis on the remaining atoms. The MLCCSD, CCSD-in-HF and CCSD/MM spectra are shifted so that the most intense peak is aligned with the corresponding CCSD peak. Lorentzian broadening with \rev{$\SI{0.3}{\eV}$} FWHM has been applied.}}
    \label{fig:methylamine_N_edge}
\end{figure}

\rev{In order to assess the quality of the MLCCSD and CCSD-in-HF approaches compared to the QM/MM approach with electrostatic embedding, we consider the nitrogen K-edge spectrum of methylamine and eight water molecules (see Figure \ref{fig:methylamine_system}) The results are given in Figure \ref{fig:methylamine_N_edge}. The active region, and the QM region of the QM/MM calculation, is methylamine. The MLCCSD, CCSD-in-HF and CCSD/MM spectra are shifted such that the most intense peak is aligned with the CCSD spectrum. Contrary to the CCSD/MM spectrum, the MLCCSD and CCSD-in-HF spectra capture the features of the CCSD spectrum. MLCCSD performs better than CCSD-in-HF, but not significantly.
}

\section{Concluding remarks}
We have presented an implementation of EOM-MLCCSD oscillator strengths for the two level CCS/CCSD model. The model can be used to simulate UV/Visible and NEXAFS spectroscopies with CCSD quality at a significantly reduced cost, given that an adequate active orbital space is employed.
In this paper, we have partitioned the orbital space by using occupied Cholesky orbitals and PAOs \rev{localized in a subregion of the molecular system}. This orbital selection procedure is suitable for localized excitation processes such as core excitations.
The CCS/CCSD model, and a reduced space CCSD model (CCSD-in-HF), has been applied to the nitrogen K-edge spectra of adenosine, adenosine triphosphate (ATP), and an ATP-water system. 
With these calculations, we have demonstrated that MLCCSD and CCSD-in-HF are useful for the accurate modeling of the NEXAFS spectra of complex molecular systems. 
\rev{Our results indicate that CCSD-in-HF may be the preferable approach to treat low lying core excitations, as the method has lower costs at no significant loss of accuracy. The MLCCSD approach, or a combined MLCCSD-in-HF approach, may be preferable for calculations with smaller active spaces and for more delocalized core excitation processes. Furthermore, MLCCSD can be used to assess the quality of CCSD-in-HF calculations.}
We believe that \rev{both} models, together with a geometry sampling from e.g. a molecular dynamics simulation, will provide a useful theoretical tool for the interpretation of experimental NEXAFS spectra of solvents and liquids.

\section{Acknowledgements}
We thank Alexander C. Paul \rev{and Tommaso Giovannini} for enlightening discussions.
We acknowledge computing resources through UNINETT Sigma2 - 
the National Infrastructure for High Performance Computing and Data Storage in Norway,
through project number NN2962k. 
We acknowledge funding from the Marie Sk{\l}odowska-Curie European Training Network ``COSINE - COmputational Spectroscopy In Natural sciences and Engineering'', Grant Agreement No. 765739 and the Research Council of Norway through FRINATEK projects 263110 and 275506.

\bibliography{main.bib}
\begin{figure}
    \centering
    \includegraphics{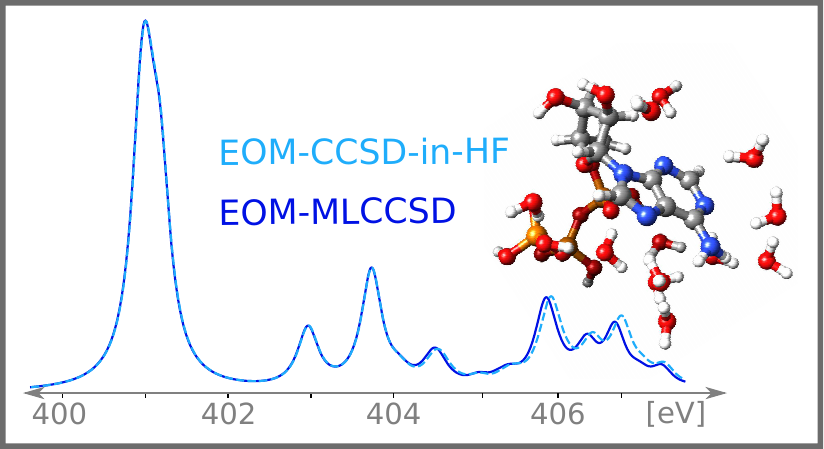}
    \caption{For Table of Contents Only}
\end{figure}
\end{document}